**Evidence of the kinetic-driven separation of butane isomers in UiO-66 (Zr) MOF, a combined experimental and computational study**

Alexander E. Khudozhitkov,[ab] Sergei S. Arzumanov,[ab] Daniil I. Kolokolov,*[ab] and Alexander G. Stepanov*[ab]

[a.] *Boreskov Institute of Catalysis, Siberian Branch of Russian Academy of Sciences, Prospekt Akademika Lavrentieva 5, Novosibirsk 630090, Russia.*

[b.] *Novosibirsk State University, Pirogova Street 2, Novosibirsk 630090, Russia*

**Abstract.**

UiO-66 is a well-known metal organic framework that proved to be promising for various separation applications. In this report we study the molecular mobility and transport of n-butane and isobutane through the UiO-66. We show that n-butane propagates through the material significantly faster than isobutane both in terms of energetics (diffusion barrier is 25 kJ mol$^{-1}$ for n-butane and 50 kJ mol$^{-1}$ for isobutane) and rate ($D_i$(303 K)/$D_n$(303 K) = 1700). This result allows expecting very high kinetic driven separation selectivity of n-butane/isobutane mixture in UiO-66.

**Introduction.**

Energy efficient separation of light linear/branched alkanes mixtures is an important technological challenge. Owing to the low boiling points, the separation of such basic compound as butane isomers becomes too energy-demanding when traditional methods such as cryogenic distillation are applied [1]. In contrast, separation methods based on shape-selective microporous molecular sieves offer an energy-efficient way to split these gas mixtures in pure components. Driven by this challenge, in the past years a lot of effort was put into developing kinetic-driven separation processes, like the microporous membrane technologies [2]. Traditionally, the main component of such membranes consisted of zeolites, for example, the MFI type ones [3]. Despite being efficient, in past decade a new class of microporous solids, the metal-organic frameworks (MOFs), emerged as a prominent platform for hydrocarbons separations [4]. Owing to their hybrid structure composed of inorganic nodes bridged by organic linkers, the MOFs are unique materials in their tunability to a specific functionality. An easier synthesis, outstanding sorption properties, the availability of some MOFs in a porous glass phase determine the deep interest into using MOFs a sorbent and molecular sieves, particularly in the field of kinetic separation in alkane/alkenes

The challenge behind the MOFs is their generally lower stability and overall complexity of adsorption and dynamical processes occurring within these highly flexible frameworks. While the latter challenge can be regarded rather as an additional degree of freedom in the MOFs functionality, the stability problem can be a real issue: within vast number (>70000 [5]) of currently known MOF structures only a small fraction are stable enough for real industrial application. In this regard, a growing attention recently received the Zr-based UiO-66 MOF compound, known for its outstanding thermal and chemical stability [6]. UiO-66 (Zr) consists of zirconium clusters connected by 1,4-benzodicarboxylic acid struts that build up a 3D porous network formed be tetrahedral (~ 8 Å) and octahedral (~ 11 Å) cages interconnected by ~ 6 Å narrow windows. Such pore-window-pore topology offers and interesting platform for adsorption-based applications. Indeed UiO-66 was positively tested for a number of catalytic [7] and separation applications,[8][9] including the challenging isobutane isomers separation. Although in all these cases the phenomena heavily depend on the mobility of the reagents, only in very few case studies [10] the guests dynamics was actually measured. Similarly little is known about the structural dynamics of the framework itself as a function of guests loading [6, 11]. Hence, we have a still typical situation for the MOF-based materials, when the molecular mechanism behind the process remains largely unknown and uncharacterized.

In this report we focus on the detailed investigation of the n-butane/isobutane isomers mobility within the UiO-66 (Zr) microporous framework in its dehydroxylated form, with the aim to shed light on the separation mechanism of these compounds on the molecular scale.

So far, the mobility and the diffusion mechanism of isobutane have not been studied at all. As for the n-butane, the diffusivity in the hydroxylated form of UiO-66 has been studied by IR spectroscopy [12]. It was shown to form weak hydrogen bond with the OH-group of inorganic node. Judging on the loss of intensity of bands associated with the bound n-butane the rate of diffusion and energetics have been derived ($E_a$=21 kJ/mol). In the dehydroxylated form of UiO-66 the mobility of n-butane has been studied by the molecular dynamics simulation [8]. At low loadings n-butane was shown to occupy preferentially tetrahedral cages. The authors associated it with stronger guest-host interaction inside the small cage. The diffusion barrier obtained by this method differs from the barrier in the hydroxylated form and is equal to 11.3 kJ/mol.

By means of $^2$H NMR spectroscopy we unravel the complex dynamics of both butane isomers in UiO-66. This method is well-known as a robust tool to probe mobility mechanism in porous solids [13]. All motional modes exhibited by these adsorbates inside the framework have been thoroughly characterized. We show that the activation barrier of diffusion in case of isobutane is twice higher than for n-butane (50 kJ/mol against 25 kJ/mol). Such drastic difference in activation barriers leads to the kinetic sieving of n-butane/isobutane mixture.

## 2. Experimental section

### 2.1 Materials

The synthesis and activation were performed according to the previously reported procedure. Butane isomers n-butane-$d_{10}$ and isobutane-$d_1$ were purchased from Sigma-Aldrich, Inc. and used further without additional purification.

### 2.2 $^2$H NMR experiments

$^2$H NMR experiments were performed at the Larmor frequency $\omega_0/2\pi$ = 61.424 MHz on a Bruker Avance-400 spectrometer using a high-power probe with a 5 mm horizontal solenoid coil. All $^2$H NMR spectra were obtained by a Fourier transform of a quadrature-detected and phase-cycled quadrupole echo after two phase-alternating 90°-pulses in the pulse sequence ($90°_x - \tau - 90°_y - \tau -$ acquisition$-$ t), where $\tau$ = 20 μs and t is a repetition time of the sequence during accumulation of the NMR signal. The duration of π/2 pulse was 1.8 μs. Spectra were typically obtained with 4−100 scans with a repetition time ranging from 0.5 to 10 s. Inversion−recovery experiments for measurements of spin−lattice relaxation times ($T_1$) were carried out using the pulse sequence $180°_x - t_v - 90°_x - \tau - 90°_y - \tau -$ acquisition $- t$ where $t_v$ was the variable delay

between the 180° and the 90°-pulses. Spin-spin relaxation time ($T_2$) was obtained by a Carr-Purcell-Meiboom-Gill pulse sequence. The repetition time *t* was always longer than 5-fold of the obtained relaxation time $T_1$.

The temperature of the samples was controlled with a flow of nitrogen gas by a variable-temperature unit BVT-3000 with a precision of about 1 K. The sample was allowed to equilibrate at least 15 min at the temperature of the experiment before the NMR signal was acquired. Modeling of spin relaxation time was performed with a homemade FORTRAN program based on the standard formalism.

### 2.3. $^2$H NMR Sample Preparation

The preparation of the sample for NMR experiments was performed in the following manner. The powder of UiO-66 (~0.07 g) was placed into a special glass cell of 5 mm diameter and 3 cm length. Then, the cell was connected to the vacuum line and activated at 473 K for 12 h under vacuum. After cooling the sample back to room temperature, the material was exposed to the vapor of the previously degassed deuterated guest molecules in the calibrated volume (55 cm$^3$). The residual quantity of the guest not consumed by the material was finally frozen with liquid nitrogen on the sample. The quantity of the adsorbed guest (1 mol per cavity = 12 mol/u.c.) was regulated by the vapor pressure inside the calibrated volume. After adsorption, the neck of the tube was sealed off, while the material sample was maintained in liquid nitrogen to prevent its heating by the flame. Prior to NMR investigations, all sealed samples were kept at 373 K for 48 h to allow an even redistribution of the guest molecules inside the porous material.

### 3. Results and Discussion

#### 3.1 Spectra lineshape analysis

Figure 1a,b shows low temperature spectra of butane isomers in UiO-66. The spectrum of isobutane is constituted by narrow Lorentzian signal with full width at half height (FWHM) 12 kHz and wider anisotropic Pake doublet signal with effective quadrupole coupling constant (QCC) $Q_{eff}$ = 84 kHz and $\eta_{eff}$ = 0.46. The spectrum of n-butane consists of Lorentzian signal with FWHM 9 kHz and two anisotropic Pake doublets with effective QCC $Q_{eff1}$ = 24 - 28 kHz, $\eta_1 = 0$ and $Q_{eff2}$ = 38 - 45 kHz, $\eta_2 = 0$ that correspond to CD$_3$ and CD$_2$ groups correspondingly (Figure 1). Presence of two spectral components implies that there are mobile guest molecules that can rotate isotropically and give Lorentzian spectrum (dynamic state I) and bound guest molecules that exhibit hindered dynamics that lead to anisotropic pattern (dynamic state II). The rate of exchange between dynamical states has to be lower than the spectrum width (ca. 50 kHz) otherwise the individual spectral components related to different dynamical states will be unresolved.

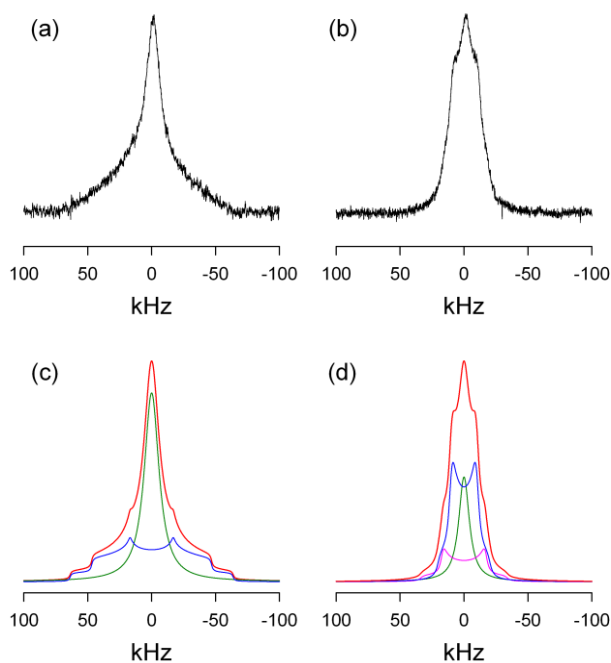

**Figure 1** $^2$H NMR spectra of (a) isobutane-$d_1$ and (b) n-butane-$d_{10}$ at 113 K adsorbed in UiO-66 (Zr). Deconvolution of experimental spectra is shown in (c) and (d).

Increase of the temperature leads to the gradual increase of narrow signal intensity and disappearance of wide signal component as illustrated for n-butane in Figure 2 and for isobutane in Figure 3. This effect can be associated with the redistribution of guest molecules between dynamical states. From the deconvolution of the spectrum lineshape we can assess the equilibrium constant for the exchange between dynamical states $K_{eq} = p_I/p_{II}$. Figure 4 shows van't Hoff plot for this equilibrium constant and allows assessing the enthalpy of the transition from the dynamical state II to the dynamical state I for n-butane $\Delta H \approx 3$ kJ mol$^{-1}$ and for isobutane $\Delta H \approx 5.4$ kJ mol$^{-1}$. Although dynamical state II thermodynamically is more preferable above 168 K both butane isomers tend to occupy dynamical state I.

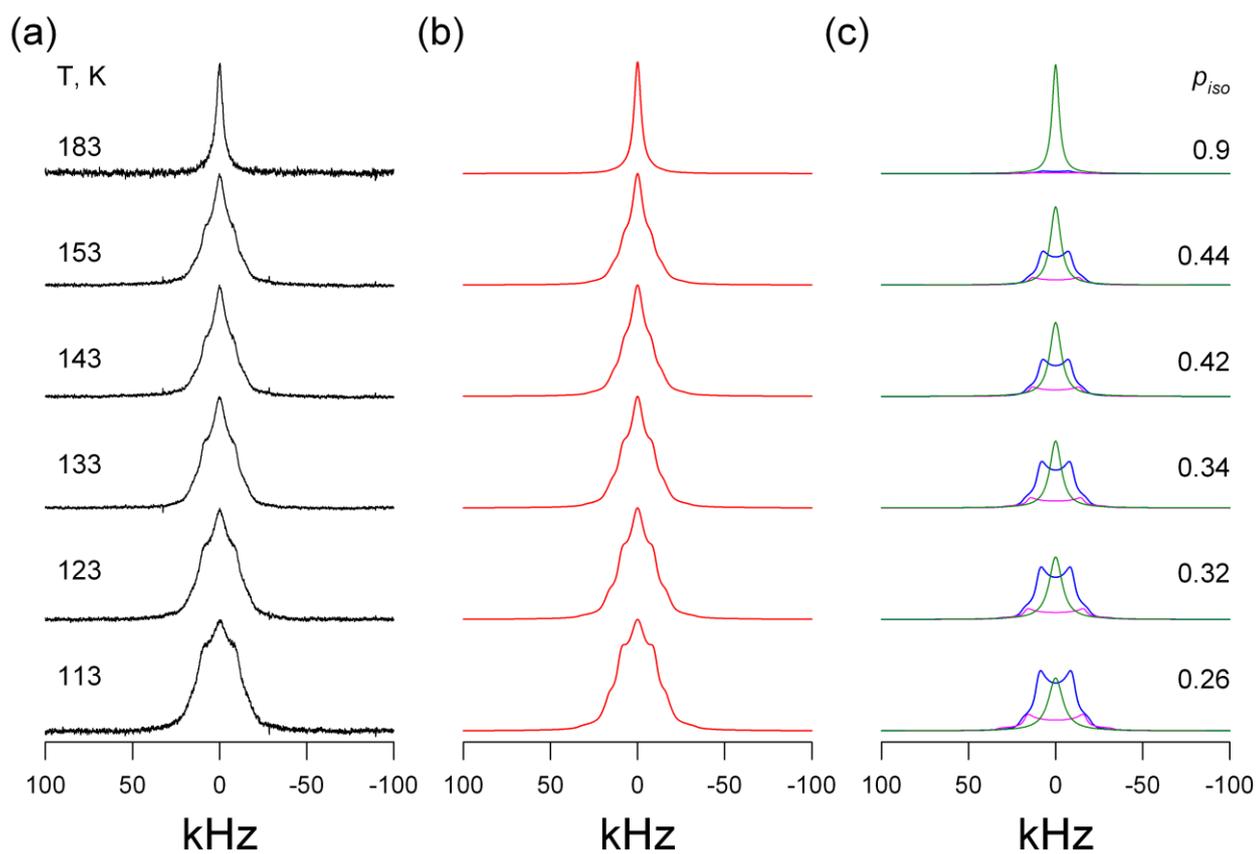

**Figure 2** $^2$H NMR spectra of n-butane adsorbed in UiO-66 (Zr). (a) experimental spectra, (b) simulated spectra and (c) deconvolution of simulated spectra.

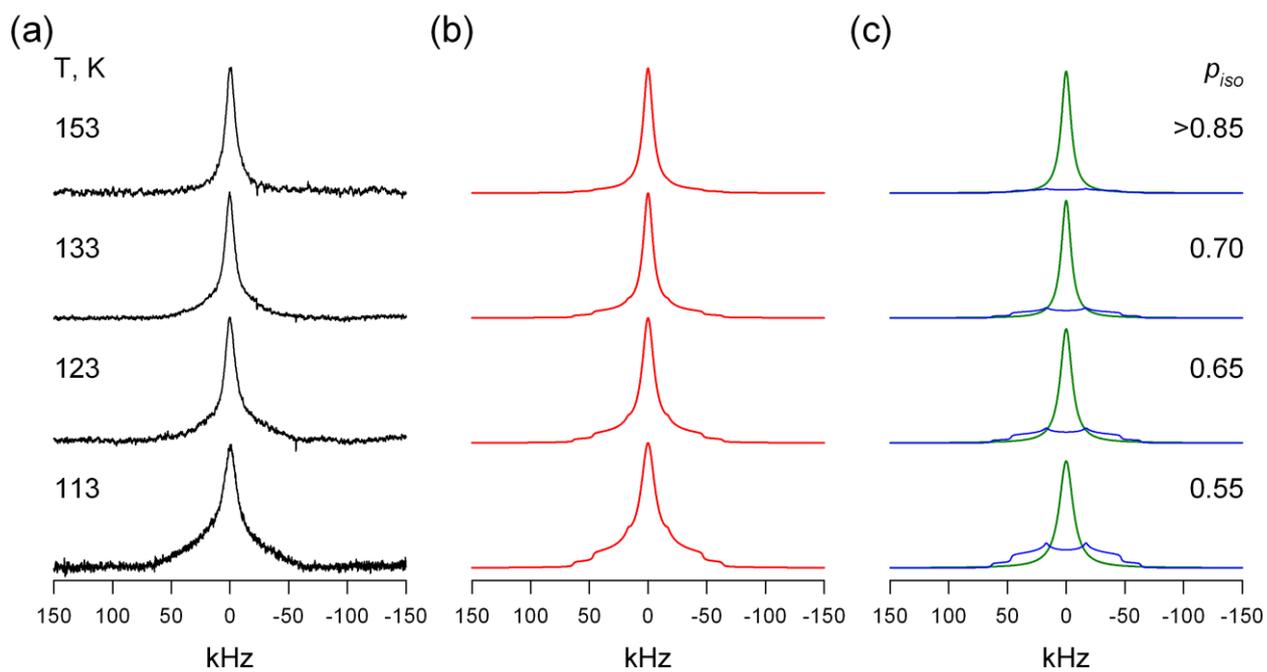

**Figure 3** $^2$H NMR spectra of isobutane adsorbed in UiO-66 (Zr). (a) experimental spectra, (b) simulated spectra and (c) deconvolution of simulated spectra.

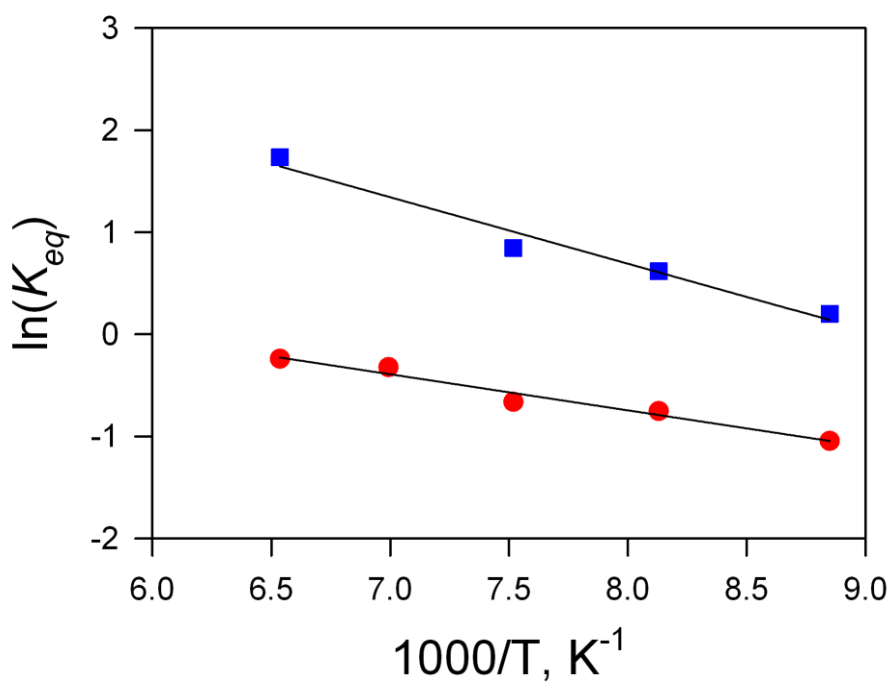

**Figure 4** Van 't Hoff graph for equilibrium constant $K_{eq} = p_I/p_{II}$ of n-butane (●) and isobutane (■) transition from tetrahedral to octahedral cage.

One possible explanation for the existence of two dynamically different species of adsorbate is provided by the topology of the UiO-66 structure. The structure of UiO-66 is comprised of tetrahedral and octahedral cages; one could expect that molecules localized in the tetrahedral cages would move in a different way than molecules in the octahedral cages. Narrow Lorentzian signal originates from the guest molecules that isotropically rotate inside the cavity with a rate higher than quadrupole coupling constant ($Q_0 \approx 176$ kHz). Since octahedral cages have larger size it is reasonable to ascribe narrow signal to molecules localized in the octahedral cages (Fig. 5a). In that case the second signal will be due to the molecules in tetrahedral cages. Jumping between cages provides the exchange between these populations.

The other possibility is inspired by the recent study of molecular mobility in ZIF-8 [14]. It was shown that even inside one cavity there may be two species of molecules with notably different motional behavior. Since the size of ZIF-8 cavity 11.6 Å is close to the size of octahedral cage ~11 Å we may expect the similar motional behavior. In that case the liquid-like signal originates from the mobile molecules localized in the canter of the cavity, whereas the anisotropic signal corresponds to the immobile molecules bound to the cage walls (Fig. 5b).

How can we distinguish these two possibilities? MD simulation shows that tetrahedral cages are occupied prior to the octahedral cages. So if we accept that the majority of molecules are localized in the tetrahedral cage than the coexistence of two populations inside one cavity seems unlikely, since tetrahedral cages are too small for that and octahedral cages are not sufficiently occupied.

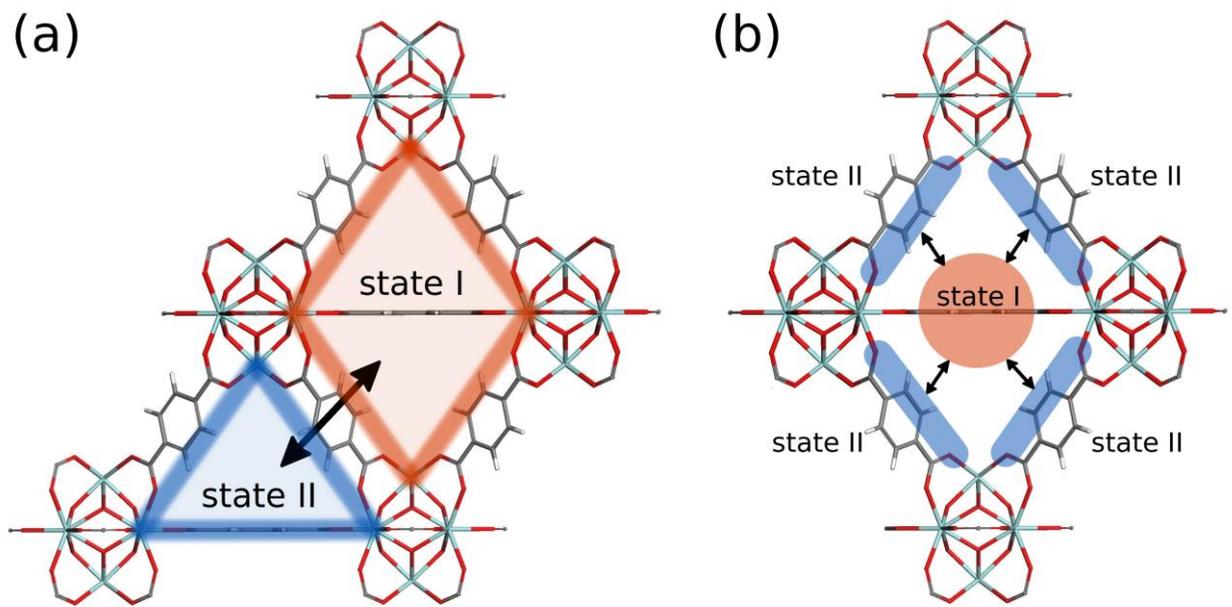

**Figure 5.** Schematic representation of possible localization of dynamical states.

### 3.2 Spin relaxation analysis

In order to determine the rates of motions we measured the rates of spin relaxation as a function of the temperature. Figure 6a shows the relaxation curves of n-butane. Both relaxation curves exhibit well resolved characteristic minima: $T_1$ relaxation has minimum at 143 K, $T_2$ relaxation is fastest at 180 K. Each of these minima corresponds to some distinct motional mode. Following the observations made on the basis of line shape analysis, the modeling of the relaxation time for n-butane takes into account two dynamical states. The relaxation times may be affected by the slow exchange between the two dynamically different states I and II.

To treat such a case the following modeling scheme was applied:[14b] we first compute the individual ($T_1^I$, $T_1^{II}$) relaxation times for the states I and II. Then we introduce the exchange rates $k_{12}$ and $k_{21}$ and the corresponding equilibrium constant $K_{eq}$ required to compute the relative populations $p_{II}$ and $p_I$. Here $k_{21} = k_{12} \times K_{eq}$, and $K_{eq} = p_I/p_{II}$, $k_{12} = k_0^{ex} \times \exp(-E_{ex}/RT)$. The temperature dependence of $K_{eq}$ is assumed to be defined by Gibbs equation: $\Delta H - T\Delta S = -RT ln K_{eq}$, where $\Delta H$ and $\Delta S$ are the enthalpy and the entropy of the transformation reaction of the state I to the state II. Having defined the exchange rates, the effective relaxation times $T_1$ can be computed directly from the Bloch's equations as the corresponding Eigen values of the respective exchange matrices.[15] Details of computation are given in the SI.

The dynamical model used to compute the individual relaxation times $T_1^I$ and $T_1^{II}$ was chosen in the following way. Dynamical state I includes the rotation of methyl group with the rate $k_{methyl}$ and isotropic rotation with the rate $k_{iso}$ (Figure 6b). We ascribe dynamical sate I to the molecules that occupy octahedral cages since it has enough space to allow isotropic tumbling of n-butane molecule. Dynamical state II includes the rotation of methyl group with the rate $k_{methyl}$, librations *in cone* of the molecular axis with the rate $k_{cone}$ and additional 2-site jumps with the rate $k_D$ that takes into account reorientation exhibited by guest molecule when it tries to squeeze through the window (Figure 6c). The dynamical state II is localized in the tetrahedral cages. The fitting parameters for the developed dynamical model are all summarized in the Table 1.

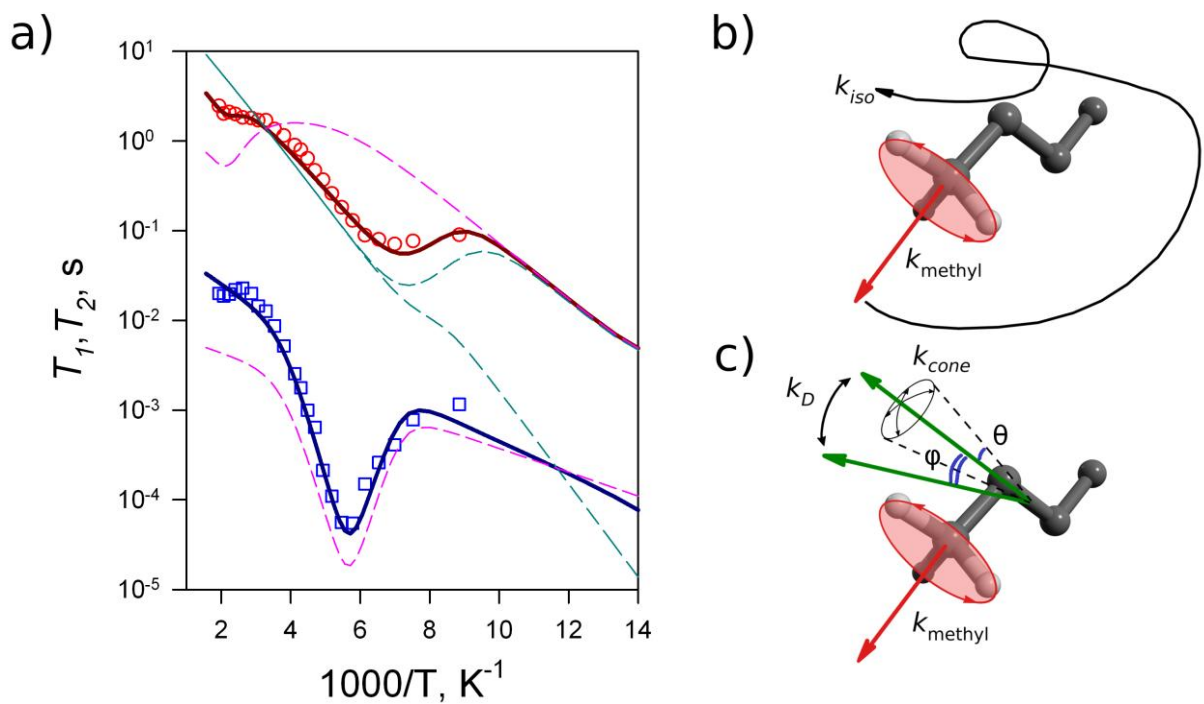

**Figure 6.** (a) Experimental $T_1$ (○) and $T_2$ (□) spin relaxation times of n-butane-$d_{10}$ adsorbed in UiO-66 (Zr). Dashed lines represent the individual relaxation times of molecules in state I (cyan lines) and state II (pink lines). Solid lines show final simulated spin relaxation times. Schematic representation of dynamic model in state I and II are shown in (b) and (c) correspondingly. All hydrogens except one methyl group are omitted for clarity.

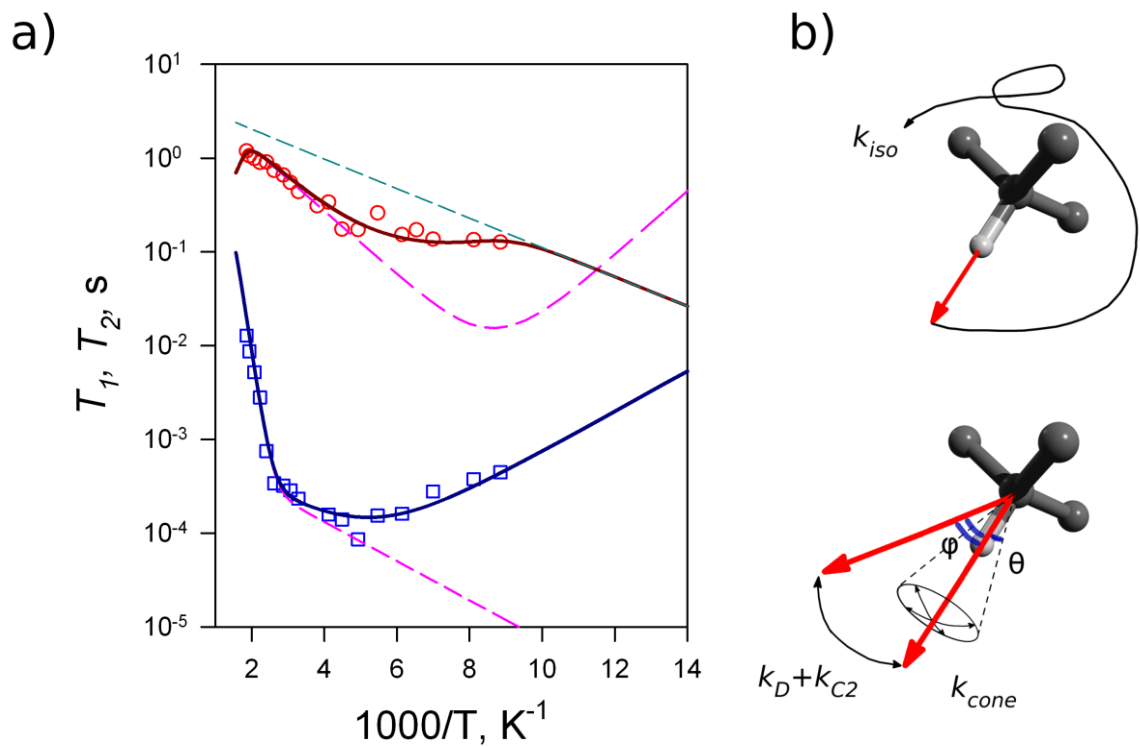

**Figure 7.** Experimental $T_1$ (○) and $T_2$ (□) spin relaxation times of isobutane-$d_1$ adsorbed in UiO-66 (Zr). Dashed lines represent the individual relaxation times of molecules confined in tetrahedral (pink lines) and octahedral (cyan lines) cages. Solid lines show final simulated spin

relaxation times. Schematic representation of dynamic model in state I and II are shown in (b) and (c) correspondingly. All hydrogens except one are omitted for clarity.

In case of isobutane relaxation curves do not have pronounced minima. $T_1$ relaxation steadily grows with the temperature, whereas $T_2$ relaxation has shallow minimum at 203 K followed by a very steep growth after 382 K (Figure 7a). Such sharp increase of $T_2$ implies high-energetic slow motion that we ascribe to diffusion. The dynamical model used to compute the individual relaxation times of isobutane-$d_1$ is similar to the model of n-butane motion except the rotation of methyl groups is not taken into account. Dynamical state I includes only the isotropic rotation with the rate $k_{iso}$ (Figure 7b). Dynamical state II includes the librations *in cone* of the molecular axis with the rate $k_{cone}$ and additional 2-site jumps characterized by two independent rate constants $k_D$ and $k_{C2}$ (Figure 7c). The fitting parameters used to reproduce experimental spin relaxation curves are presented in the Table 1.

**Table 1** Motional parameters of butane isomers used in the simulation of spin relaxation curves.

|  | isobutane | n-butane |
|---|---|---|
| $E_{methyl}$, kJ mol$^{-1}$ | – | 6 |
| $k_0^{methyl}$, s$^{-1}$ | – | $8 \times 10^{12}$ |
| $E_{C2}$, kJ mol$^{-1}$ | 4(3-6) | – |
| $k_0^{C2}$, s$^{-1}$ | (4-15) $6 \times 10^5$ | – |
| $E_{cone}$, kJ mol$^{-1}$ | 6.5(5-7) | 2.5(2-3.5) |
| $k_0^{cone}$, s$^{-1}$ | (5-10) $9 \times 10^{10}$ | (3.5-5) $5 \times 10^6$ |
| $E_{iso}$, kJ mol$^{-1}$ | 3 (2-8) | 10 (10-11.5) |
| $k_0^{iso}$, s$^{-1}$ | (1-600) $3 \times 10^{11}$ | (5-20) $7 \times 10^{11}$ |
| $E_D$, kJ mol$^{-1}$ | (45-65) 50 | 25 |
| $k_0^D$, s$^{-1}$ | (1-160) $6 \times 10^{11}$ | $5 \times 10^{10}$ |
| $\Delta H$, kJ mol$^{-1}$ | – 8 (7-10) | 3 |
| $E_{ex}$, kJ mol$^{-1}$ | 1.5 | 1.5 |
| $k_0^{ex}$, s$^{-1}$ | $10^7$ | $10^7$ |

The estimated accuracy is 10% for all activation barriers and 20% for all pre-exponential factors.

Comparison of the motional parameters for butane isomers provides us the following observations:

- Both isomers have slow motional mode ($k_0 \sim 10^6$ s$^{-1}$) and relatively low activation barrier (2.5 – 4 kJ mol$^{-1}$)
- Isotropic rotation of isobutane has lower activation barrier than n-butane (3 vs 10 kJ mol$^{-1}$)

- Motion ascribed to diffusion has drastically higher activation barrier for isobutane and becomes detectable at temperature higher by 200 K than for n-butane

The value of $\Delta H$ obtained from spin relaxation analysis for n-butane is in a good agreement with the spectrum lineshape behavior.

On the first sight the lower value of activation barrier for isotropic rotation in case of isobutane looks strange since its kinetic diameter $d_i = 5.3$ Å is larger than the diameter of n-butane $d_n = 4.3$ Å. However, the maximal dimension of n-butane exceeds the diameter of almost spherical isobutane, so apparently the confinement of octahedral cage is tight enough to be more sensitive to this property rather than to kinetic diameter.

The activation barrier of motion that we ascribe to diffusion is twice higher for isobutane than for n-butane (50 kJ mol$^{-1}$ vs. 25 kJ mol$^{-1}$). The elongated shape of n-butane makes its minimal dimension smaller than that of isobutane facilitating the jumping between cages. The activation barrier for isobutane measured in this work is significantly higher than the value 11.3 kJ mol$^{-1}$ obtained from MD calculation in dehydroxylated form [8]. Surprisingly, it is closer to the results of Sharp et al. that studied the diffusion of n-butane in hydroxylated form of UiO-66 [12]. The activation barrier of n-butane diffusion in hydroxylated form is 21 kJ mol$^{-1}$ that is close to 25 kJ mol$^{-1}$ measured in this work. Moreover, we can assess diffusion coefficient at infinite temperature $D_0$ according to formula $D_0 = \langle l^2 \rangle / 6\tau$, where $l \approx 1$ Å is a mean jump length between cages and $\tau = 1/(2\pi k_0^D)$ is a characteristic time between jumps. The value $D_0 = 5\times10^{-4}$ cm$^2$ s$^{-1}$ derived in this way is almost identical with the value $D_0 = 10^{-4}$ cm$^2$ s$^{-1}$ provided by Sharp et al. Since the activation temperature 473 K guarantees full dehydroxylation of the sample, we assume that the diffusion of n-butane is similar in both hydroxylated and dehydroxylated form. However, additional study of the hydroxylated UiO-66 is required to support this hypothesis.

Slow motional mode that is present for both butane isomers has the same order of magnitude as the motional mode of various guests adsorbed in ZIF-8 [14]. Following our previous work we assume that this motion is related to some collective vibrational modes of the framework that guest molecule exhibits in the vicinity of the cavity walls.

### 4. Conclusion

A thorough characterization of n-butane and isobutane mobility in UiO-66 has been performed by $^2$H NMR method. Two populations with drastically different motional behavior have been found: highly mobile state I and bound state II. The localization of these dynamic states is not clear yet and requires additional investigation. Highly energetic motions that we ascribe to diffusion show that intercage transport is faster for n-butane. The activation barrier of n-butane diffusion 25 kJ mol$^{-1}$ is twice lower than the barrier of isobutane diffusion 50 kJ mol$^{-1}$. Moreover, the diffusivities at room temperature ($2\times10^{-11}$ cm$^2$ s$^{-1}$ for isobutane and $3.4\times10^{-8}$ for n-butane) differ by three orders of magnitude providing theoretically outstanding selectivity of butane isomer separation. The smaller minimal dimension of n-butane leads to the faster transport of this isomer between cavities.

**AUTHOR INFORMATION**

**Corresponding Authors**

*E-mail: kdi@ catalysis.ru (D.I.K.).

*E-mail: stepanov@catalysis.ru (A.G.S.); Tel : +7 9529059559 ; Fax: +7 383 330 8056

**Author Contributions**

The manuscript was written through contributions of all authors.

**Notes**

The authors declare no competing financial interests.

**ACKNOWLEDGMENT**

The reported study was funded by RFBR, project number 19-33-90026.